\documentclass[12pt,letterpaper]{article}
\usepackage{amsmath,amssymb,array,calc,rotating,epsfig,psfrag,amscd, cite}

\makeatletter
\renewcommand\section{\@startsection {section}{1}{\z@}%
                                   {-3.5ex \@plus -1ex \@minus -.2ex}%nn
                                   {2.3ex \@plus.2ex}%
                                   {\normalfont\large\bfseries}}
\renewcommand\subsection{\@startsection{subsection}{2}{\z@}%
                                     {-3.25ex\@plus -1ex \@minus -.2ex}%
                                     {1.5ex \@plus .2ex}%
                                     {\normalfont\bfseries}}
\makeatother

\let\non\nonumber

\let\s=\sigma

\let\S=\Sigma

\newcommand{\bea}{\begin{eqnarray}}
\newcommand{\eea}{\end{eqnarray}}
\newcommand{\be}{\begin{equation}}
\newcommand{\ee}{\end{equation}}

\newcommand{\p}{\partial}

\newcommand{\C}[1]{$(\ref{#1})$}

\def\IZ{\relax\ifmmode\mathchoice
{\hbox{\cmss Z\kern-.4em Z}}{\hbox{\cmss Z\kern-.4em Z}}
{\lower.9pt\hbox{\cmsss Z\kern-.4em Z}} {\lower1.2pt\hbox{\cmsss
Z\kern-.4em Z}}\else{\cmss Z\kern-.4em Z}\fi}
\def\IR{\relax{\rm I\kern-.18em R}}

\def\one{{\hbox{ 1\kern-.8mm l}}}

\newlength{\bredde}
\def\slash#1{\settowidth{\bredde}{$#1$}\ifmmode\,\raisebox{.15ex}{/}
\hspace*{-\bredde} #1\else$\,\raisebox{.15ex}{/}\hspace*{-\bredde}
#1$\fi}

\newsavebox{\zzzbar}
\sbox{\zzzbar}
  {\setlength{\unitlength}{0.9em}
  \begin{picture}(0.6,0.7)
  \thinlines
  \put(0,0){\line(1,0){0.6}}
  \put(0,0.75){\line(1,0){0.575}}
  \multiput(0,0)(0.0125,0.025){30}{\rule{0.3pt}{0.3pt}}
  \multiput(0.2,0)(0.0125,0.025){30}{\rule{0.3pt}{0.3pt}}
  \put(0,0.75){\line(0,-1){0.15}}
  \put(0.015,0.75){\line(0,-1){0.1}}
  \put(0.03,0.75){\line(0,-1){0.075}}
  \put(0.045,0.75){\line(0,-1){0.05}}
  \put(0.05,0.75){\line(0,-1){0.025}}
  \put(0.6,0){\line(0,1){0.15}}
  \put(0.585,0){\line(0,1){0.1}}
  \put(0.57,0){\line(0,1){0.075}}
  \put(0.555,0){\line(0,1){0.05}}
  \put(0.55,0){\line(0,1){0.025}}
  \end{picture}}

\newcommand{\ena}{\end{eqnarray}}
\newcommand{\beqa}{\begin{eqnarray}}
\newcommand{\eeqa}{\end{eqnarray}}

\def\s{\sigma}

\def\S{\Sigma}

\begin{document}
\begin{titlepage}

\begin{center}

%\hfill \today
%\hfill         \phantom{xxx}         

%\hfill HRI

\vskip 2 cm
{\Large \bf Simplifying the one loop five graviton amplitude in type IIB string theory}\\
\vskip 1.25 cm { Anirban Basu$^{a,b}$\footnote{email address:
    anirbanbasu@hri.res.in} } \\
{\vskip 0.5cm $^a$ Harish--Chandra Research Institute, Chhatnag Road, Jhusi,\\
Allahabad 211019, India\\$^b$   Homi Bhabha National Institute, Training School Complex, \\ Anushakti Nagar,
     Mumbai 400085, India\\}

\end{center}

\vskip 2 cm

\begin{abstract}
\baselineskip=18pt

We consider the $D^8\mathcal{R}^5$ and $D^{10}\mathcal{R}^5$ terms in the low momentum expansion of the five graviton amplitude in type IIB string theory at one loop. They involve integrals of various modular graph functions over the fundamental domain of $SL(2,\mathbb{Z})$. Unlike the graphs which arise in the four graviton amplitude or at lower orders in the momentum expansion of the five graviton amplitude where the links are given by scalar Green functions, there are several graphs for the $D^8\mathcal{R}^5$ and $D^{10}\mathcal{R}^5$ terms where two of the links are each given by a derivative of the Green function.
Starting with appropriate auxiliary diagrams, we show that these graphs can be expressed in terms of those which do not involve any derivatives. This results in considerable simplification of the amplitude.

\end{abstract}

\end{titlepage}

%\pagestyle{plain}
%\baselineskip=18pt
% Try a wider skip
%\baselineskip=19pt
%%%%%%%%%%%%%%%%%%%%%%%%%%%%%%%%%%%%%%%%%%%%%%%%%%%%%%%%%%%%%%%%%%%%%%%%%%%%%%

\section{Introduction}

The effective action of string theory contains invaluable information about the perturbative and non--perturbative states of the theory. The moduli dependent coefficients of the various interactions in the effective action are covariant under duality symmetries of the theory, which when expanded around weak coupling yields information about S--matrix elements in perturbative string theory. Thus it is useful to understand these perturbative amplitudes in detail, which also sometimes yield powerful insight into the non--perturbative structure of these amplitudes by imposing constraints of duality covariance.        

Our aim is to consider certain one loop amplitudes in type IIB string theory. These yield local interactions in the effective action on performing the low momentum expansion. The total contribution at a fixed order in the $\alpha'$ expansion reduces to an integral of the form  
\be \label{F}\sum_i \int_{\mathcal{F}_L} \frac{d^2\tau}{\tau_2^2} f_i(\tau,\bar\tau)\ee     
where $\tau$ is the complex structure modulus of the torus, and $d^2\tau = d\tau_1 d\tau_2$. The overall momentum dependence of the amplitude has been factored out. The modular graph functions $f_i (\tau,\bar\tau)$ involve integrals over the insertion points of the various vertex operators on the toroidal worldsheet. The vertices of these graphs are the positions of insertions of the vertex operators, while the links represent Green functions (possibly with derivatives) connecting the vertices. The sum over $i$ runs over the topologically distinct graphs.  The integral in \C{F} is over the truncated fundamental domain of $SL(2,\mathbb{Z})$ defined by~\cite{Green:1999pv}
\be \label{one}\mathcal{F}_L = \Big\{ -\frac{1}{2} \leq \tau_1 \leq \frac{1}{2}, \vert \tau \vert \geq 1, \tau_2 \leq L\Big\},\ee  
where $L \rightarrow \infty$, which produces finite as well as contributions that diverge as $L\rightarrow \infty$. The finite part is the local contribution to the amplitude, while the divergent part cancels that from the boundary of moduli space that is obtained from integrating over $L \leq \tau_2 \leq \infty$. The finite part of the later contribution yields the non--analytic terms in the effective action. These non--analytic contributions are obtained by analyzing the asymptotic nature of the integrand and is non--perturbative in the external momenta.        

Thus in order to calculate the one loop amplitude, it is very useful to understand the structure of these modular graph functions. They have been been analyzed in detail for the four graviton amplitude for the first few terms in the low momentum expansion~\cite{D'Hoker:2015foa,D'Hoker:2015zfa,Basu:2015ayg,D'Hoker:2015qmf,D'Hoker:2016jac,Basu:2016xrt,Basu:2016kli}. Poisson equations have been derived for them which allow us to perform the integrals over moduli space and obtain the amplitude~\cite{D'Hoker:2015foa,Basu:2015ayg,Basu:2016fpd}. What is particularly significant about the structure of these graphs is that they satisfy various relations among each other. Hence topologically distinct graphs can be expressed in terms of a far lesser number of them, which simplifies the calculation of these amplitudes. As an aside, they also answer the question of how many of them are really independent, which is interesting to study in its own right.      

For the four graviton amplitude, the graphs that arise involve the scalar Green functions that connect the various vertices, but not their derivatives. However, the Poisson equations satisfied by them can involve source terms given by graphs that have derivatives of Green functions. This is indeed the case for the Poisson equation for the three loop ladder graph function~\cite{Basu:2016xrt}, which has a source term given by a graph involving two links each of which has a derivative of the Green function, which arises in the five graviton amplitude\footnote{Note that the number of vertices is not fixed in the Poisson equation.}. In fact, topologically distinct graphs involving derivatives of Green functions arise for higher point graviton amplitudes in the type II theory~\cite{Green:2013bza}.  

Thus the graphs which involve derivatives raises several interesting questions. One would like to know if they can be expressed in terms of graphs with no derivatives, in which case the structure of the amplitude simplifies considerably since the Poisson equations for such graphs can be obtained along the lines of~\cite{D'Hoker:2016jac,Basu:2016kli}. If not, it is useful to know how many of them are independent, and what Poison equations do they satisfy. Our aim is to address these issues for the graphs that arise in the low momentum expansion of the five graviton amplitude. While this provides the simplest setting for addressing these issues, explicit calculations can be done given the details of the amplitude.         

To be specific, we consider the low momentum expansion of the type IIB five graviton amplitude. For terms upto $O(D^6\mathcal{R}^5)$ in the expansion, there are no graphs which involve derivatives. For terms yielding the $D^8\mathcal{R}^5$ and $D^{10} \mathcal{R}^5$ interactions, while there are several graphs with links involving only Green functions, there are two and seven topologically distinct graphs respectively, which involve two links each of which has a derivative of the Green function~\cite{Green:2013bza}. In order to attempt to express them in terms of graphs without derivatives, we do not find it particularly useful to deal with these diagrams directly. Instead, we introduce appropriate auxiliary diagrams~\cite{Basu:2016xrt,Basu:2016kli} with more links and derivatives, which are related to the graphs we are interested in. Manipulating them directly, we are able to express all the graphs involving derivatives in terms of graphs with no derivatives of Green functions. Apart from a few new graphs without derivatives, most of the graphs without derivatives are already present in the expressions for the one loop five graviton amplitude, and this additional contribution merely changes the coefficients. We analyze these terms in the low energy expansion of the amplitude along the lines of~\cite{D'Hoker:2016jac,Basu:2016kli} to obtain their contribution in ten dimensions. Our result leads to considerable simplification of the amplitude as one does not have to consider graphs with derivatives at all. It would be interesting to generalize the analysis to higher orders in the momentum expansion and also to higher point functions where there are more involved diagrams involving derivatives of Green functions to see how much of this structure survives. 

We begin by considering the analysis of the two diagrams for the $D^8\mathcal{R}^5$ interaction, which is followed by the analysis of the seven diagrams for the $D^{10} \mathcal{R}^5$ interaction. We then simplify the resulting expressions that we obtain on removing the derivatives. We also analyze the $D^6\mathcal{R}^5$ interaction for the sake of completeness. Relevant formulae are given in the appendix. 
      
\section{The various modular graph functions}

We first consider the various modular graph functions that are relevant to our analysis~\cite{D'Hoker:2015foa,D'Hoker:2016jac,Basu:2016kli}, where the conventions for the various diagrams are mentioned in the appendix. The relevant graph functions with two and three links are given in figure 1. They only involve Green functions. We have that
\bea F_1 = E_2, \quad F_2 = E_3, \quad F_3 = E_3 +\zeta(3),\eea
where the Eisenstein series $E_s$ is defined in \C{Eisenstein}. Thus these graphs are given by simple expressions.

\begin{figure}[ht]
\begin{center}
\[
\mbox{\begin{picture}(220,100)(0,0)
\includegraphics[scale=.9]{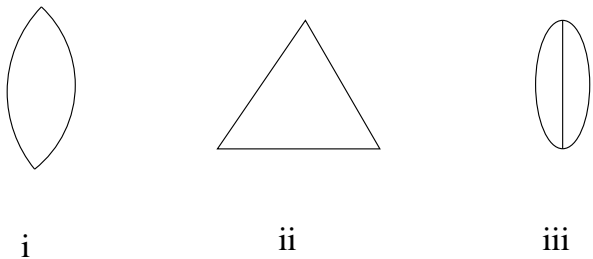}
\end{picture}}
\]
\caption{The graphs (i) $F_1$, (ii) $F_2$ and (iii) $F_3$}
\end{center}
\end{figure}

We next consider the relevant graphs with five links given in figure 2, which only involve Green functions. Among them, we have that
\be F_4 = E_5, \quad F_5 = \frac{2}{5}E_5 +\frac{\zeta(5)}{30}\ee
and hence are given by simple expressions. The remaining graphs satisfy Poisson equations, and hence are more involved.

\begin{figure}[ht]
\begin{center}
\[
\mbox{\begin{picture}(240,200)(0,0)
\includegraphics[scale=.65]{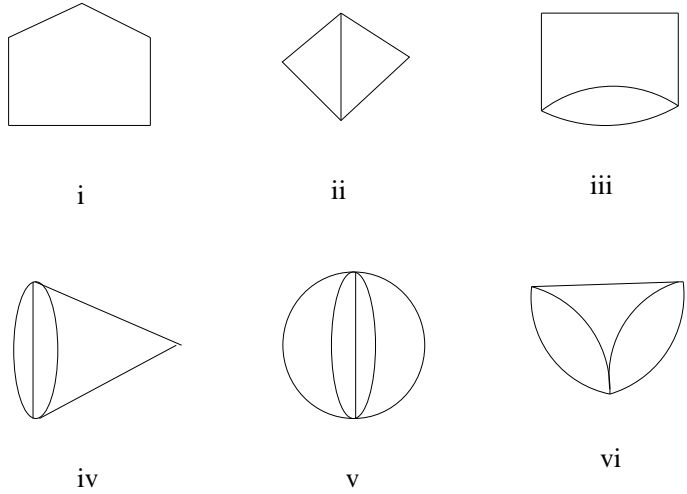}
\end{picture}}
\]
\caption{The graphs (i) $F_4$, (ii) $F_5$, (iii) $F_6$, (iv) $F_7$, (v) $F_8$ and (vi) $F_9$}
\end{center}
\end{figure}

We now consider the graph functions that involve derivatives of Green functions~\cite{Green:2013bza}. For the $D^8\mathcal{R}^5$ term, there are two diagrams $B_1$ and $B_2$ given in figure 3. $B_1$ is manifestly real, while the reality of $B_2$ follows from the analysis below. Hence they directly arise in the expression for the five graviton amplitude.   

For the $D^{10}\mathcal{R}^5$ term, there are seven diagrams $B_3$, $B_4$, $B_5$, $B_6$, $B_7$, $B_8$ and $B_9$ given in figure 4. While the reality of $B_3$, $B_4$, $B_5$, $B_6$ and $B_9$ is manifest, the reality of $B_8$ follows from the analysis below. It is the combination $B_7 +c.c.$ that arises in the expression for the five graviton amplitude.

\begin{figure}[ht]
\begin{center}
\[
\mbox{\begin{picture}(220,100)(0,0)
\includegraphics[scale=.75]{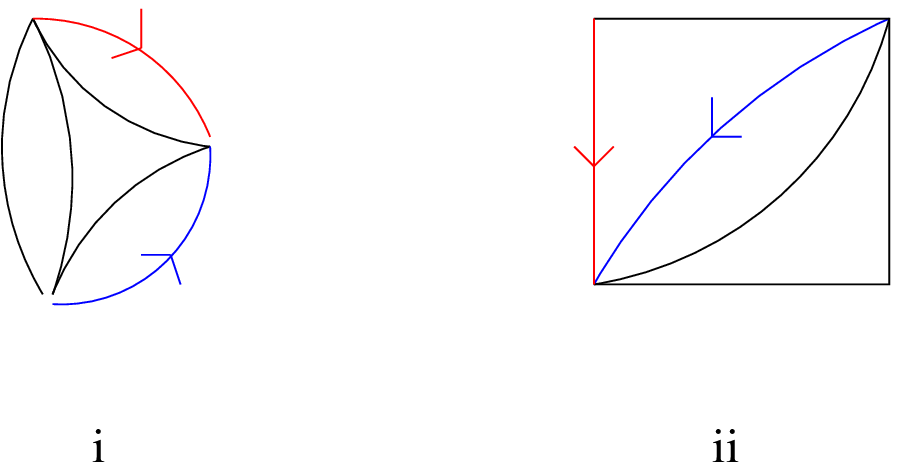}
\end{picture}}
\]
\caption{The graphs (i) $B_1$ and (ii) $B_2$}
\end{center}
\end{figure}

\begin{figure}[ht]
\begin{center}
\[
\mbox{\begin{picture}(380,160)(0,0)
\includegraphics[scale=.75]{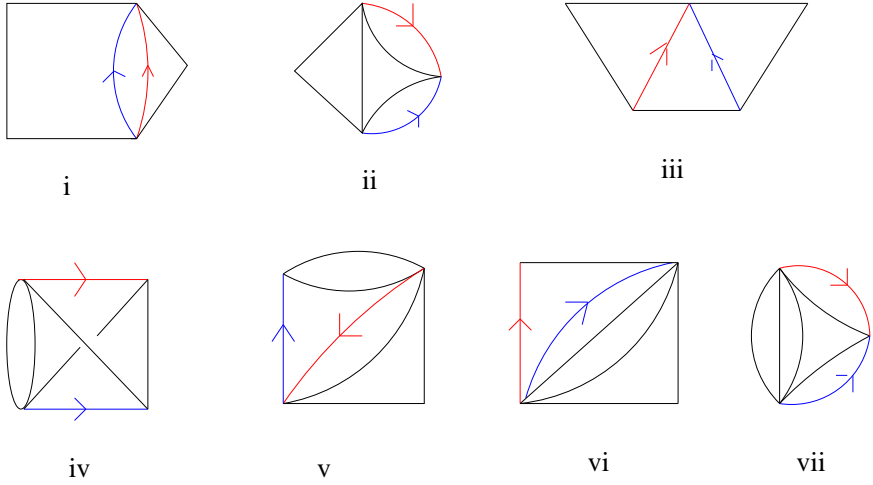}
\end{picture}}
\]
\caption{The graphs (i) $B_3$, (ii) $B_4$, (iii) $B_5$, (iv) $B_6$, (v) $B_7$, (vi) $B_8$ and (vii) $B_9$}
\end{center}
\end{figure}

The nine graph functions given in figures 3 and 4 are the central ones in our analysis. In our analysis below we ignore various expressions that vanish identically.

\section{The graphs with derivatives for the $D^8\mathcal{R}^5$ term}

In this section we consider the two graphs with derivatives of Green functions given in figure 3 for the $D^8\mathcal{R}^5$ term.

\subsection{The analysis for $B_1$}

To begin with, we have that
\be \label{one1}B_1 = -\frac{1}{2} C_1 - \frac{\pi}{2} F_9 \ee
where $C_1$ is given in figure 5.

\begin{figure}[ht]
\begin{center}
\[
\mbox{\begin{picture}(180,110)(0,0)
\includegraphics[scale=.6]{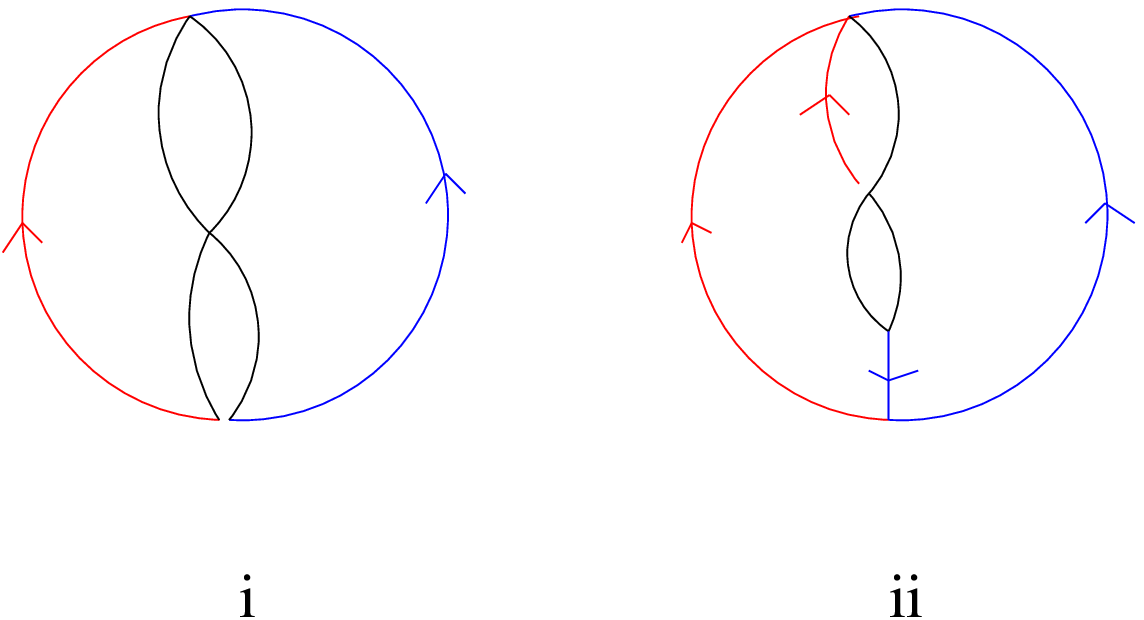}
\end{picture}}
\]
\caption{ The graphs (i) $C_1$ and (ii) $C_2$}
\end{center}
\end{figure}

In obtaining the expression \C{one1}, we have dropped a contribution that involves a Green function $G(z,z)$ which forms a closed loop~\cite{Green:2013bza}. This is the regularization which is followed throughout the analysis later on as well. This is because such contributions have to be dealt with separately when the vertex operators collide on the worldsheet. Their operator product expansion leads to the massless poles in the amplitude, and hence they do not lead to contact terms in the effective action. These contributions have to be calculated by letting $z_i$ get close to $z_j$ directly in the integrand where these are the positions of insertions of the colliding vertex operators and extracting the contribution. Note that one does not perform the $\alpha'$ expansion of the integrand while separating the pole contribution. We neglect such contributions for our purposes.

Now to calculate $C_1$, we start with the auxiliary diagram $C_2$ given in figure 5 leading to
\be C_2 = \frac{\pi}{2} C_1.\ee

Next to calculate $C_2$ we start with the auxiliary diagram $C_3$ given in figure 6. 
\begin{figure}[ht]
\begin{center}
\[
\mbox{\begin{picture}(100,90)(0,0)
\includegraphics[scale=.65]{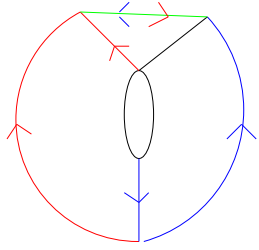}
\end{picture}}
\]
\caption{The graph $C_3$}
\end{center}
\end{figure}

We evaluate this graph in two ways. One is the trivial way using the equation~\C{eigen} for the $\p$ and $\bar\p$ on the same link. This gives us
\be C_3 = \pi C_2 -\pi^2 C_4 +\pi^2 C_5 -\pi^3 F_9 + \pi^3 E_2 E_3\ee 
where $C_4$ and $C_5$ are given in figure 7.

\begin{figure}[ht]
\begin{center}
\[
\mbox{\begin{picture}(160,110)(0,0)
\includegraphics[scale=.6]{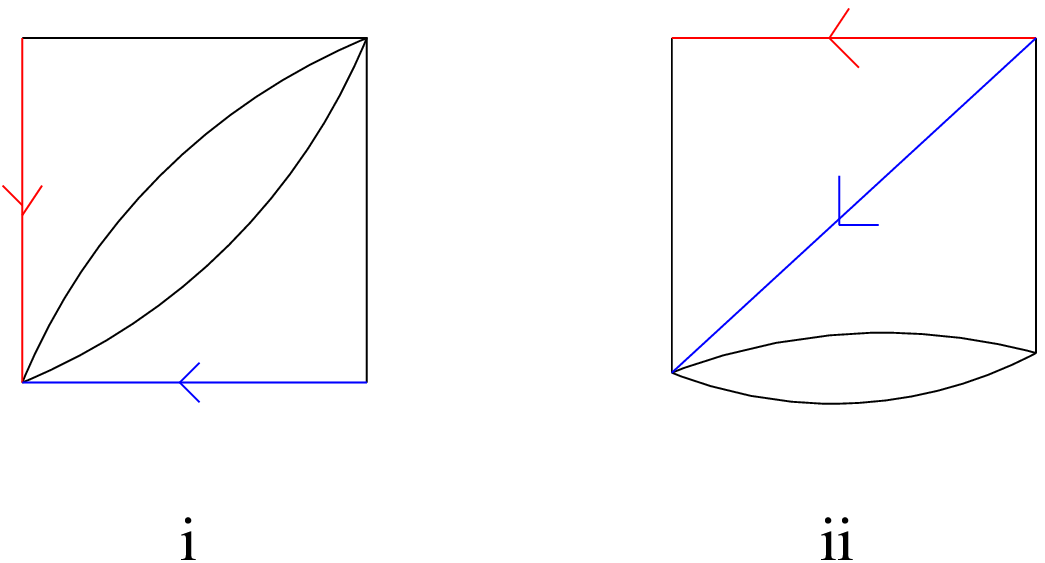}
\end{picture}}
\]
\caption{ The graphs (i) $C_4$ and (ii) $C_5$}
\end{center}
\end{figure}

We alternatively evaluate $C_3$ by moving the various arrows along the various links using momentum conservation leading to
\be C_3 = 2\pi^2 C_5 +\pi^2 C_5^* -\pi^3 F_6  -\frac{\pi^3}{3} F_8 - \frac{\pi^3}{2} F_9 +\frac{\pi^3}{3} E_2 F_3,\ee
which gives us
\be B_1 = - C_4 -(C_5 + c.c.)   +\pi F_6 +\frac{\pi}{3} F_8- \pi F_9- \frac{\pi}{3} E_2 F_3 +\pi E_2 E_3.\ee

Using the relations~\cite{Basu:2016kli}
\bea \label{e}\pi^{-1}C_4 &=& 4 F_5 + 3 F_6 +F_7- 2 E_2 E_3 - E_2 F_3 ,\non \\ \pi^{-1} (C_5 +c.c.)&=& -F_6 -F_7 + F_9  + E_2 E_3 +E_2 F_3, \eea
we get that
\be \label{B1}\pi^{-1} B_1 = -4 F_5- F_6  +\frac{1}{3} F_8 -2 F_9 +2 E_2 E_3-\frac{1}{3} E_2 F_3 .\ee

\subsection{The analysis for $B_2$}

Next we have that
\be B_2 = \frac{\pi}{2} F_7 - \frac{1}{2} C_4,\ee
where $C_4$ is given in figure 7, hence $B_2$ is real.

Now using the expression for $C_4$ in \C{e} we get that

\be \label{B2}\pi^{-1}B_2 =   -2 F_5 - \frac{3}{2} F_6+E_2 E_3 +\frac{1}{2} E_2 F_3.\ee
 
Thus from \C{B1} and \C{B2} we see that both the graphs have been expressed in terms of graphs with no derivatives of Green functions. 

\section{The graphs with derivatives for the $D^{10}\mathcal{R}^5$ term}

We first mention the new graphs with four links that are needed in our analysis. They are given in figure 8, where $F_{10} = E_4$. The others satisfy Poisson equations.  

\begin{figure}[ht]
\begin{center}
\[
\mbox{\begin{picture}(200,100)(0,0)
\includegraphics[scale=.6]{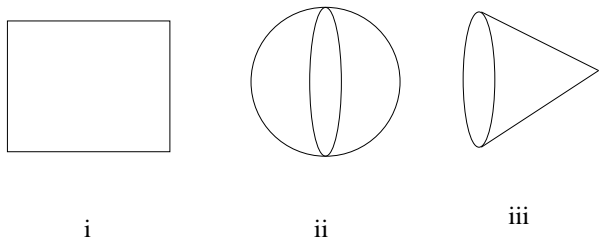}
\end{picture}}
\]
\caption{The graphs (i) $F_{10}$, (ii) $F_{11}$ and (iii) $F_{12}$}
\end{center}
\end{figure}

We next consider the various graphs with six links without any derivatives that will be needed in our analysis. They are given in figure 9, where $F_{13} = E_6$. The others are not all independent, as will be discussed later.

\begin{figure}[ht]
\begin{center}
\[
\mbox{\begin{picture}(290,210)(0,0)
\includegraphics[scale=.6]{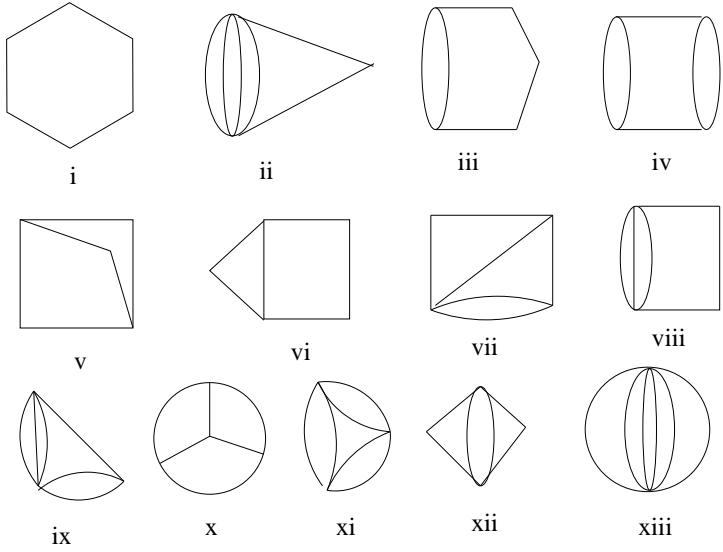}
\end{picture}}
\]
\caption{The graphs(i) $F_{13}$, (ii) $F_{14}$, (iii) $F_{15}$, (iv) $F_{16}$, (v) $F_{17}$, (vi) $F_{18}$, (vii) $F_{19}$, (viii) $F_{20}$, (ix) $F_{21}$ , (x) $F_{22}$, (xi) $F_{23}$, (xii) $F_{24}$ and (xiii) $F_{25}$}
\end{center}
\end{figure}

We also impose the equality given in figure 10 in our analysis~\cite{Basu:2016xrt}.
 
\begin{figure}[ht]
\begin{center}
\[
\mbox{\begin{picture}(120,65)(0,0)
\includegraphics[scale=.7]{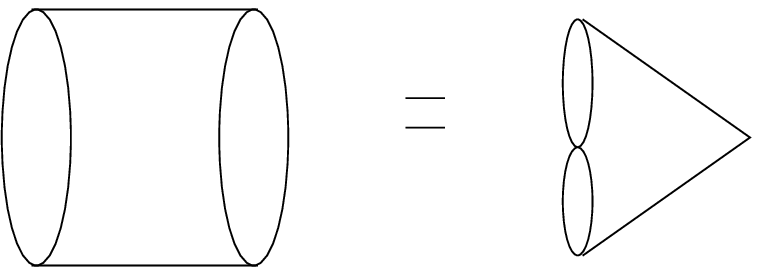}
\end{picture}}
\]
\caption{An equality among graphs}
\end{center}
\end{figure}

\subsection{The analysis for $B_3$}

To begin with, we have that
\be \label{Rel}2 B_3 = (F_{26} + c.c.)-2\pi F_{18} - \pi F_{20} -\pi F_{24} ,\ee
where $F_{26}$ is given in figure 11.

\begin{figure}[ht]
\begin{center}
\[
\mbox{\begin{picture}(420,80)(0,0)
\includegraphics[scale=.7]{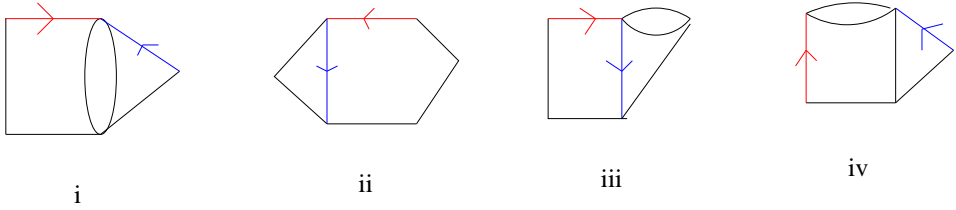}
\end{picture}}
\]
\caption{The graphs (i) $F_{26}$, (ii) $F_{29}$, (iii) $F_{30}$ and (iv) $F_{31}$}
\end{center}
\end{figure}

To evaluate it, we start with the auxiliary diagram $F_{27}$ given in figure 12, leading to
\be F_{27} = \frac{\pi}{2} F_{26} - \frac{\pi^2}{2} E_2 E_4. \ee

\begin{figure}[ht]
\begin{center}
\[
\mbox{\begin{picture}(170,100)(0,0)
\includegraphics[scale=.7]{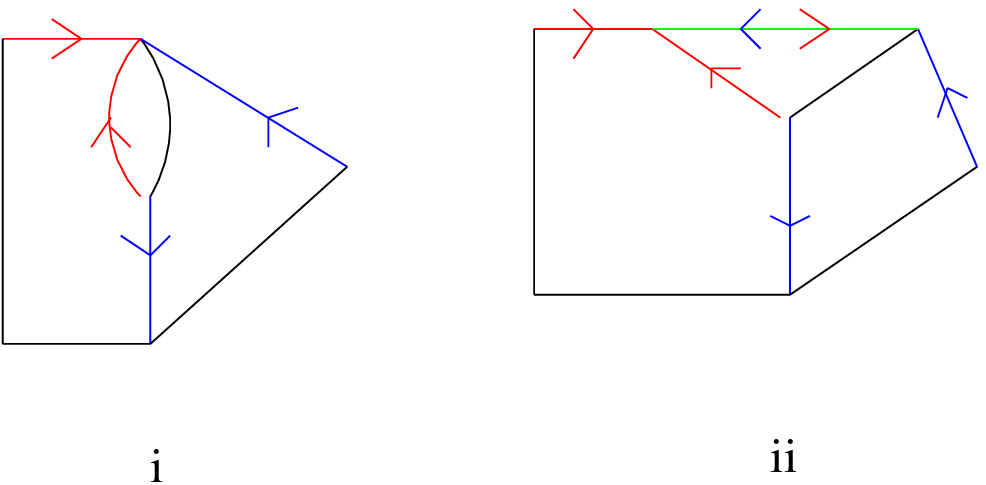}
\end{picture}}
\]
\caption{The graphs (i) $F_{27}$ and (ii) $F_{28}$}
\end{center}
\end{figure}

To evaluate $F_{27}$, we consider the auxiliary diagram $F_{28}$ instead given in figure 12. 

Evaluating it trivially and otherwise, we get that
\bea F_{28} &=& \pi F_{27} +\pi^2 F_{29} +\pi^3 E_6 \non \\ &=& -\frac{\pi^3}{2} F_{16} +\frac{\pi^3}{2} F_{17}  +2\pi^3 F_{18} +\frac{\pi^3}{2} F_{19}+2\pi^2 F_{29} +\frac{\pi^2}{2} F_{30} +\frac{\pi^2}{2} F_{31},\eea
where $F_{29}, F_{30}$ and $F_{31}$ are given in figure 11. 

Then using the relations
\bea \label{Rel2}\pi^{-1} F_{29} +c.c. &=&  F_{15}-F_{18}  + E_6 - E_2 E_4, \non \\ \pi^{-1} F_{30} + c.c. &=& F_{15}- F_{19}+F_{20}   - E_2 E_4 -E_3 F_3, \non \\ \pi^{-1} F_{31} + c.c. &=& F_{15} + F_{16}+F_{19}  -E_2 F_{12} ,\eea
we get that
\be \label{B3}\pi^{-1} B_3 =  2 F_{15} - \frac{1}{2} F_{16}+ F_{17} +F_{19}   + 2 F_{18} -\frac{1}{2} F_{24} -E_6 -\frac{1}{2} E_2 E_4 -\frac{1}{2} E_3 F_3 -\frac{1}{2} E_2 F_{12}.\ee

\subsection{The analysis for $B_4$}

To analyze $B_4$, we start with the auxiliary diagram $F_{32}$ instead given in figure 13 which gives us
\be F_{32} = \pi B_4 +\frac{\pi^2}{6} F_{17} +\pi^2 F_{19} -\frac{\pi^2}{2} F_{24}   +\pi^2 E_3^2 - \frac{\pi^2}{6} E_2^3.\ee

\begin{figure}[ht]
\begin{center}
\[
\mbox{\begin{picture}(220,110)(0,0)
\includegraphics[scale=.9]{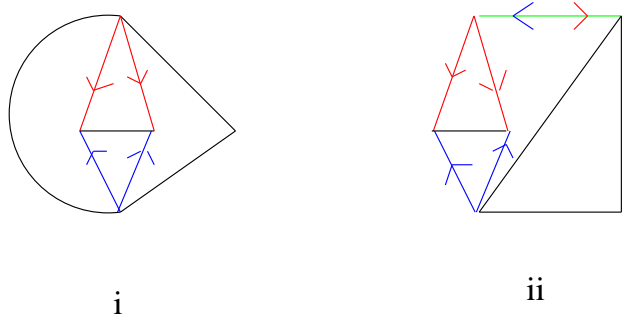}
\end{picture}}
\]
\caption{The graphs (i) $F_{32}$ and (ii) $F_{33}$}
\end{center}
\end{figure}

To calculate $F_{32}$, we consider yet another auxiliary diagram $F_{33}$ given in figure 13. Evaluating it trivially and otherwise, we get that
\bea F_{33} &=& \pi F_{32} - \pi^3 E_3^2 \non \\ &=& \frac{\pi^3}{3} F_{14} +2\pi^3 F_{18}-\pi^3 F_{19}  - \pi^3 E_2 F_{12} -\frac{\pi^3}{3} E_3 F_3.\eea 

Substituting the various expressions, we get that
\be \label{B4}\pi^{-1}B_4 = \frac{1}{3} F_{14} -\frac{1}{6} F_{17} +2F_{18} -2 F_{19}  +\frac{1}{2} F_{24}  -\frac{1}{3} E_3 F_3 - E_2 F_{12} +\frac{1}{6} E_2^3.\ee

\subsection{The analysis for $B_5$}

To analyze $B_5$, we start with the auxiliary diagram $F_{34}$ given in figure 14
 which gives us the identity
\be F_{34} = \pi B_5 = \pi^2 F_{15} - \pi^2 F_{18} +F_{35},\ee
where the diagram $F_{35}$ is given in figure 14. Thus in order to analyze $B_5$, it is enough to analyze the diagram $F_{35}$.  

\begin{figure}[ht]
\begin{center}
\[
\mbox{\begin{picture}(330,100)(0,0)
\includegraphics[scale=.7]{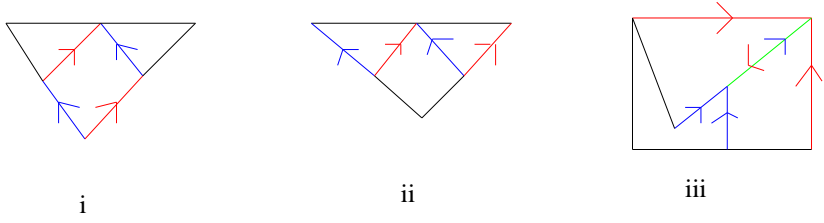}
\end{picture}}
\]
\caption{The graphs (i) $F_{34}$, (ii) $F_{35}$ and (iii) $F_{36}$}
\end{center}
\end{figure}

Instead of directly analyzing $F_{35}$, we start with the auxiliary diagram $F_{36}$ given in figure 14. Evaluating it trivially and otherwise, we get that
\bea F_{36} &=& \pi F_{35} +\pi^3 F_{18}  +\frac{\pi^3}{2} E_6 - \frac{\pi^3}{2} E_3^2, \non \\ &=&  -\frac{\pi^3}{6} F_{17} +\frac{\pi^3}{2} F_{24} +2\pi^3 E_6-\pi^3 E_2 E_4 -\pi^3 E_3^2 +\frac{\pi^3}{6} E_2^3 +\frac{\pi^2}{2} (F_{37} +c.c.),\eea 
where $F_{37}$ is given in figure 15. 

\begin{figure}[ht]
\begin{center}
\[
\mbox{\begin{picture}(60,70)(0,0)
\includegraphics[scale=.7]{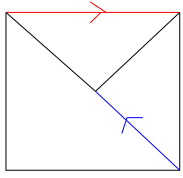}
\end{picture}}
\]
\caption{The graph $F_{37}$}
\end{center}
\end{figure}

Now using the relation
\be F_{37} + c.c. = \pi F_{16} -\pi F_{17} +2\pi F_{18} -2\pi F_{19} +\pi F_{22}, \ee
we get that
\be \label{B5}\pi^{-1} B_5 =  F_{15} +\frac{1}{2} F_{16} -\frac{2}{3} F_{17}-F_{18} -F_{19} +\frac{1}{2} F_{22} +\frac{1}{2} F_{24} +\frac{3}{2} E_6- E_2 E_4 -\frac{1}{2} E_3^2 +\frac{1}{6} E_2^3.\ee

\subsection{The analysis for $B_6$}

To analyze $B_6$, we start with the auxiliary diagram $F_{38}$ given in figure 16, which gives us that
\be F_{38} = \frac{\pi}{2} B_6 -\frac{\pi^2}{2} F_{19} +\frac{\pi^2}{2} E_2 F_{12} -\pi F_{39} +\pi F_{40} +\pi F_{41} -\frac{\pi}{2} F_{42} +\pi F_{42}^*,\ee
where $F_{39}, F_{40}, F_{41}$ and $F_{42}$ are given in figure 17. 

\begin{figure}[ht]
\begin{center}
\[
\mbox{\begin{picture}(160,110)(0,0)
\includegraphics[scale=.7]{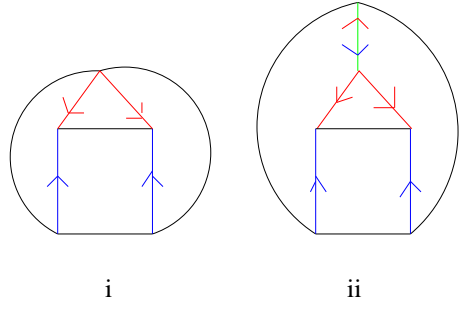}
\end{picture}}
\]
\caption{The graphs (i) $F_{38}$ and (ii) $F_{43}$}
\end{center}
\end{figure}

\begin{figure}[ht]
\begin{center}
\[
\mbox{\begin{picture}(440,50)(0,0)
\includegraphics[scale=.8]{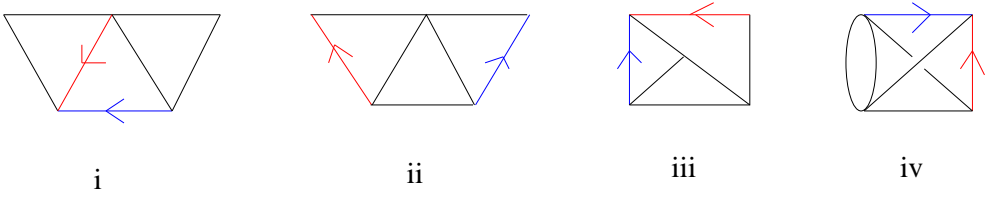}
\end{picture}}
\]
\caption{The graphs (i) $F_{39}$, (ii) $F_{40}$, (iii) $F_{41}$ and (iv) $F_{42}$}
\end{center}
\end{figure}

In order to calculate $F_{38}$ differently, we consider the auxiliary diagram $F_{43}$ in figure 16. Evaluating it trivially and otherwise, we get that
\bea F_{43} &=&  -\pi^3 F_{18} +\frac{\pi^3}{2} F_{19}-\pi^3 F_{24}+ \pi F_{38} +\pi^2 F_{39}   +\pi^3 E_3^2, \non \\ &=& -\pi^2 B_6 +2\pi^3 F_{18} -2\pi^2 F_{40} -2\pi^2 F_{41}+\pi^2 F_{42}^*  -\pi^3 E_2 F_{12}.\eea  

This yields a relation between $B_6$ and the other diagrams, which we further simplify using the reality of $B_6$.  This is done using the relations
\bea F_{40} &=& B_5 +\pi F_{18} +\pi F_{19} -\pi E_2 F_{12}, \non \\ F_{41} +c.c. &=& \pi F_{22}, \non \\ F_{42} +c.c. &=& \pi F_{23} - \pi F_{24},\eea
which gives us that
\be \label{B6}\pi^{-1} B_6 = -2 F_{19} -F_{22} +\frac{1}{6} F_{23} +\frac{1}{2}F_{24} -\frac{2}{3} E_3^2 + E_2 F_{12}-2 \pi^{-1}B_5,\ee
which is further simplified using the expression for $B_5$ in \C{B5}.

\subsection{The analysis for $B_7$}

To analyze $B_7$, we have that
\be B_7 = -\frac{\pi}{2} F_{14} +\frac{\pi}{2} E_2 F_{12} +\frac{1}{2} F_{44},\ee
where the graph $F_{44}$ is given in figure 18.

\begin{figure}[ht]
\begin{center}
\[
\mbox{\begin{picture}(90,70)(0,0)
\includegraphics[scale=.6]{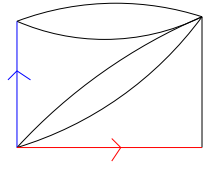}
\end{picture}}
\]
\caption{The graph $F_{44}$ }
\end{center}
\end{figure}

Now to calculate $F_{44}$, we consider the auxiliary diagram $F_{45}$ instead, given in figure 19, which gives us
\be F_{45} = \frac{\pi}{2} F_{44} - \frac{\pi^2}{2} E_2 F_{12}.\ee 

\begin{figure}[ht]
\begin{center}
\[
\mbox{\begin{picture}(200,100)(0,0)
\includegraphics[scale=.7]{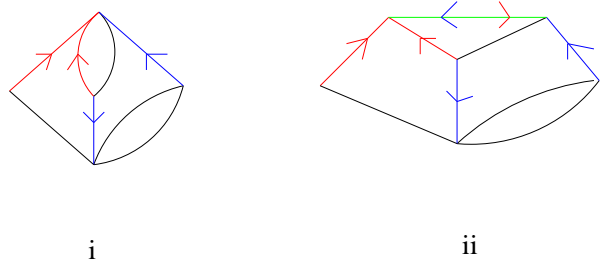}
\end{picture}}
\]
\caption{The graphs (i) $F_{45}$ and (ii) $F_{46}$ }
\end{center}
\end{figure}

To calculate $F_{45}$, we start with the auxiliary diagram $F_{46}$ instead given in figure 19. Evaluating it trivially and otherwise, we get that
\bea F_{46} &=& \pi F_{45} +\pi^3 F_{15} +\pi^2 F_{30} \non \\ &=& \pi^3 F_{19}  +\pi^2 F_{26} +\pi^2 F_{30} +\pi^2 F_{31}^* +\frac{\pi^2}{2} F_{47} +\frac{\pi^2}{2} F_{48} -\pi^3 E_2 E_4 +\frac{\pi^3}{2} E_2^3 ,\eea
where $F_{47}$ and $F_{48}$ are given in figure 20. Now $F_{48}$ is real and is given by
\be \pi^{-1} F_{48} = \frac{1}{2} F_{16} + F_{21} - E_2 F_{12} - \frac{1}{2} F_3^2.\ee

\begin{figure}[ht]
\begin{center}
\[
\mbox{\begin{picture}(160,90)(0,0)
\includegraphics[scale=.75]{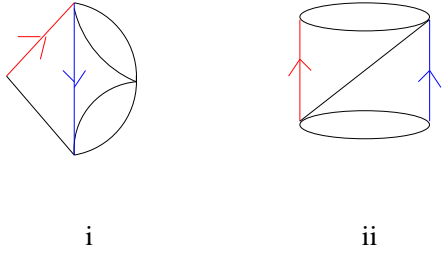}
\end{picture}}
\]
\caption{The graphs (i) $F_{47}$ and (ii) $F_{48}$ }
\end{center}
\end{figure}

Thus using the relations for $F_{26} +c.c.$ from \C{Rel}, $F_{31} +c.c.$ from \C{Rel2}, and 
\bea F_{47} + c.c. &=& 4 B_4 +\pi F_{16}-\pi F_{23} - \pi E_2 F_{11}, \eea
we get that
\bea \label{B7}\pi^{-1} B_7 + c.c. &=& -F_{14}-F_{15} +2F_{16}+2F_{18}+ 3F_{19} +F_{20} +F_{21}-\frac{1}{2}F_{23} +F_{24}  \non \\ &&+2\pi^{-1} (B_3+B_4) -2 E_2 E_4 -\frac{1}{2} E_2 F_{11} -\frac{1}{2} F_3^2 + E_2^3,\eea
which further simplifies on substituting the expressions for $B_3$ and $B_4$ given by \C{B3} and \C{B4} respectively.

\subsection{The analysis for $B_8$}

To analyze $B_8$, we have that
\be B_8 = \frac{\pi}{3} F_{14} -\frac{1}{3} F_{49}\ee
where $F_{49}$ is given in figure 21. Hence $B_8$ is real.

\begin{figure}[ht]
\begin{center}
\[
\mbox{\begin{picture}(250,90)(0,0)
\includegraphics[scale=.6]{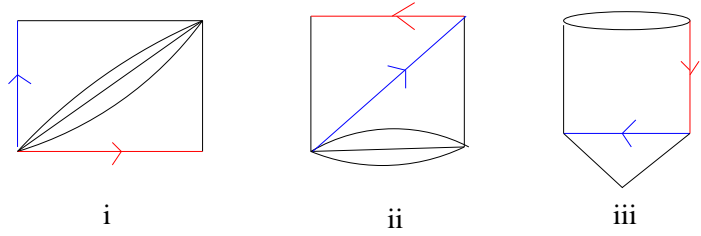}
\end{picture}}
\]
\caption{The graphs (i) $F_{49}$, (ii) $F_{52}$ and (iii) $F_{57}$}
\end{center}
\end{figure}

To calculate $F_{49}$, we consider the auxiliary diagram $F_{50}$ given in figure 22, which gives us 
\be F_{50} = \frac{\pi}{3} F_{49} -\frac{\pi^2}{3} E_3 F_3. \ee

\begin{figure}[ht]
\begin{center}
\[
\mbox{\begin{picture}(200,100)(0,0)
\includegraphics[scale=.7]{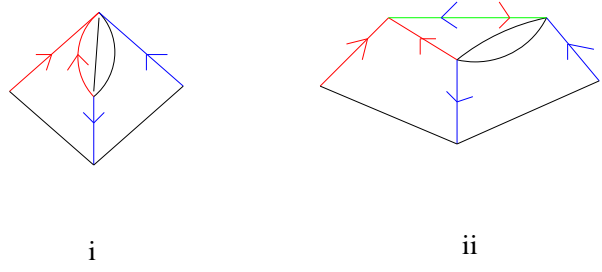}
\end{picture}}
\]
\caption{The graphs (i) $F_{50}$ and (ii) $F_{51}$}
\end{center}
\end{figure}

To calculate $F_{50}$ directly, we consider the auxiliary diagram $F_{51}$ instead given in figure 22. Evaluating it trivially and otherwise, we get that
\bea F_{51} &=& \pi F_{50} +\pi^3 F_{15} +\pi^2 F_{30}, \non \\ &=& \pi^3 F_{14} +\pi^3 F_{15} +\pi^3 F_{20} -2\pi^3 E_2 E_4 -\pi^3 E_3 F_3 -\pi^3 E_2 F_{12} +2\pi^2 B_7^* \non \\ && +\pi^2 F_{26} +\frac{\pi^2}{3} F_{52} -2\pi F_{53},\eea
where $F_{52}$ and $F_{53}$ are given in figures 21 and 23 respectively.

\begin{figure}[ht]
\begin{center}
\[
\mbox{\begin{picture}(240,110)(0,0)
\includegraphics[scale=.7]{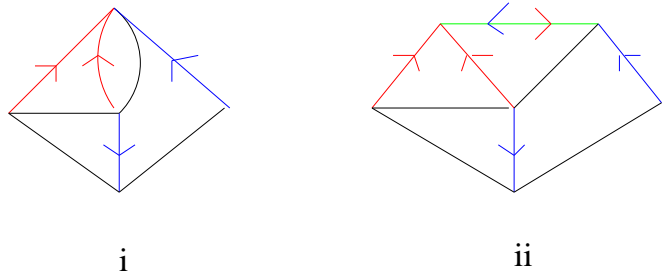}
\end{picture}}
\]
\caption{The graphs (i) $F_{53}$ and (ii) $F_{54}$}
\end{center}
\end{figure}

In order to evaluate $F_{53}$ we start with the auxiliary diagram $F_{54}$ instead given in figure 23. Evaluating it trivially and otherwise, we get that
\bea F_{54} &=& \pi F_{53} - \pi F_{55}\non\\  &=& -\frac{\pi^3}{6} F_{14} -\frac{3\pi^3}{8} F_{16} +\frac{\pi^3}{2} F_{17} -2\pi^3 F_{18} + 2\pi^3 F_{19} -\frac{\pi^3}{6} F_{20} +\frac{\pi^3}{2} F_{21} +2\pi^2B_5 \non \\ &&
-2\pi^2 F_{39}^* +\frac{\pi^2}{6} F_{52} +\frac{\pi^2}{2} F_{30}^*-\frac{\pi^2}{2} F_{31}+\frac{\pi^2}{2} F_{52}^*+\frac{\pi^3}{6} E_3 F_3 -\frac{\pi^3}{8} F_3^2+\frac{\pi^3}{6} E_2 F_{11},\non \\ \eea
where $F_{55}$ is given in figure 24. 

\begin{figure}[ht]
\begin{center}
\[
\mbox{\begin{picture}(280,100)(0,0)
\includegraphics[scale=.8]{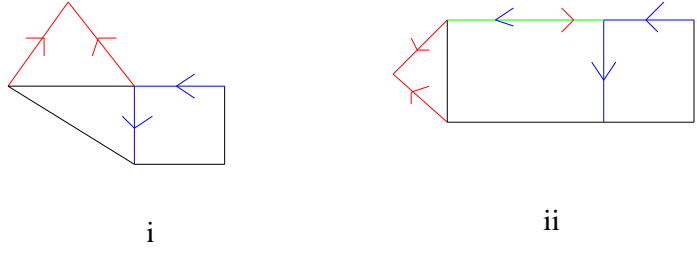}
\end{picture}}
\]
\caption{The graphs (i) $F_{55}$ and (ii) $F_{56}$}
\end{center}
\end{figure}

To calculate $F_{55}$, we start with the auxiliary diagram $F_{56}$ instead in figure 24, which evaluated trivially and otherwise leads to
\bea F_{56} &=& \pi F_{55} \\ \non &=& -\frac{\pi^3}{2} F_{15} -\pi^2 B_5 +\pi^2F_{39}^* +\frac{\pi^2}{2} F_{57}-2\pi^3 E_6+\pi^3 E_2 E_4,\eea
where $F_{57}$ is given in figure 21.

Thus substituting the various expressions and using the reality of $B_8$, we finally get that
\bea \label{B8}\pi^{-1} B_8 &=&  -4 E_6   -\frac{1}{2} F_{14} -\frac{3}{4} F_{16} +F_{17} -4 F_{18}  +3 F_{19}   - \frac{1}{3} F_{20} \non \\ &&+\frac{1}{2} F_{21} -\frac{3}{2} F_{24} - \pi^{-1} B_3  +2\pi^{-1} B_5  -\pi^{-1} (B_7+c.c.) \non \\ && + E_3^2 + 3E_2 E_4 -\frac{1}{2} E_3 F_3-\frac{1}{4} F_3^2 -\frac{1}{6} E_2 F_{11},\eea
where we have used the relations \C{Rel}, \C{Rel2} and
\bea \pi^{-1}F_{52} + c.c. &=& F_{14}+F_{20} -F_{21} - E_2 F_{11} -E_3 F_3, \non \\ \pi^{-1}F_{57} + c.c. &=&  F_{15} + F_{16} - F_{19} -E_2 F_{12}.\eea
On using the expressions \C{B3}, \C{B5} and \C{B7} for $B_3$, $B_5$ and $B_7+c.c.$ respectively in \C{B8}, this is further simplified.

\subsection{The analysis for $B_9$}

To analyze the contribution from $B_9$, we consider the auxiliary diagram $F_{58}$ given in figure 25 which gives us
\be F_{58} = \pi B_9 -3\pi B_8 +\pi^2 E_3 F_3 +\pi^2 F_{21}.\ee 

\begin{figure}[ht]
\begin{center}
\[
\mbox{\begin{picture}(220,100)(0,0)
\includegraphics[scale=.7]{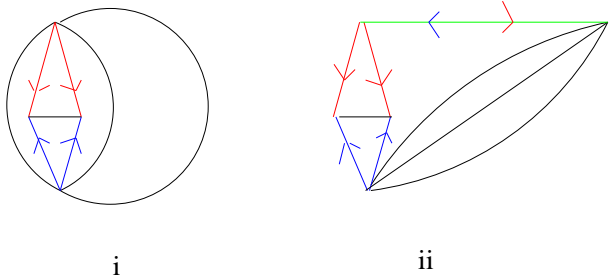}
\end{picture}}
\]
\caption{The graphs (i) $F_{58}$ and (ii) $F_{59}$}
\end{center}
\end{figure}

We now evaluate $F_{58}$ differently, starting from the auxiliary diagram $F_{59}$ in figure 25. Evaluating $F_{59}$ trivially and otherwise, we get that 
\bea F_{59} &=& \pi F_{58} - \pi^3 E_3 F_3 \non \\ &=& 2\pi^3 F_{20} -\pi^3 F_{21} + \frac{\pi^3}{3} F_{25} - \frac{\pi^3}{3} F_3^2 -\pi^3 E_2 F_{11}.\eea

Thus we get that
\be \label{B9}\pi^{-1}B_9 = 3 \pi^{-1}B_8 +2 F_{20} -2 F_{21} +\frac{1}{3} F_{25} -\frac{1}{3} F_3^2 - E_2 F_{11},\ee
which simplifies using the expression for $B_8$ in \C{B8}.

Thus from \C{B3}, \C{B4}, \C{B5}, \C{B6}, \C{B7}, \C{B8} and \C{B9} we see that all the graphs with two derivatives are expressible in terms of graphs with no derivatives. Hence upto this order in the momentum expansion, it is enough to consider only graphs with no derivatives in the five graviton amplitude.  Note that there need not be a unique set of auxiliary diagrams for a modular graph function that leads to the final expression. We simply choose ones that work.

\section{Simplifying the low momentum expansion of the five graviton amplitude}

 Using the relations we have derived that express the graphs with the derivatives of Green functions in terms of those without derivatives, we simplify their structure and evaluate them. We perform the analysis in the type IIB theory in ten dimensions keeping terms upto the $D^{10} \mathcal{R}^5$ interaction. While there are several contributions that have exactly the same structure as the four graviton amplitude  and hence can be analyzed along the lines of~\cite{D'Hoker:2016jac,Basu:2016kli}, there are certain other contributions that have a different structure, which we focus on. They are given by the modular graph functions in equation (5.6) of~\cite{Green:2013bza}. The dependence on Mandelstam invariants can be written down based on the expressions in~\cite{Green:2013bza}. We do not give the involved explicit expressions as they are not directly relevant for our purposes, our primary goal is to evaluate the integrals of these graphs over $\mathcal{F}_L$. Thus from now onwards, we consider only those expressions that are different from the four graviton ones. 

For the $D^6\mathcal{R}^5$ interaction, we have that 
\bea \mathcal{A}_{D^6\mathcal{R}^5} = -10\pi \int_{\mathcal{F}_L} \frac{d^2\tau}{\tau_2^2} \Big( \frac{F_{11}}{96}  - F_{12} - \frac{99}{80} E_4 +\frac{19}{32} E_2^2 \Big) \eea
where the graphs do not involve derivatives of Green functions. On using the relation~\cite{D'Hoker:2015zfa,D'Hoker:2016jac,Basu:2016kli}
\be F_{11} = 24 F_{12} -18 E_4 + 3 E_2^2, \ee
we get that
\be  \mathcal{A}_{D^6\mathcal{R}^5} = 10\pi \int_{\mathcal{F}_L} \frac{d^2\tau}{\tau_2^2} \Big( \frac{3}{4} F_{12} +\frac{57}{40} E_4 - \frac{5}{8} E_2^2\Big).\ee
Further using~\cite{D'Hoker:2016jac,Basu:2016kli} 
\be (\Delta -2) F_{12} = 9 E_4 - E_2^2,\ee
and the asymptotic expansions of $F_{12}$ and $E_s$ we evaluate the integral along the lines of~\cite{D'Hoker:2015foa}, to obtain
\be  \mathcal{A}_{D^6\mathcal{R}^5} =  \frac{2\zeta(2)\zeta(3)}{3} {\rm ln}\mu,\ee
where
\be {\rm ln} \mu = \frac{5}{4}- {\rm ln}2 + \frac{\zeta' (3)}{\zeta(3)} - \frac{\zeta'(4)}{\zeta(4)}.\ee

For the $D^8\mathcal{R}^5$ interaction, we have that
\bea \mathcal{A}_{D^8\mathcal{R}^5} = -\frac{24\pi}{5} \int_{\mathcal{F}_L} \frac{d^2\tau}{\tau_2^2} \Big( \frac{F_5}{4} +\frac{F_6}{8} +\frac{F_7}{6} +\frac{7}{96} F_8 -\frac{F_9}{8} + E_5 +\frac{13}{48} E_2 F_3 -\frac{B_1}{4\pi} -\frac{B_2}{\pi}\Big).\eea
On using the relations~\cite{D'Hoker:2016jac,Basu:2016kli} 
\bea 40 F_7 &=& 300 F_6 -276 E_5+120 E_2 E_3  +7\zeta(5), \non \\ F_8 &=& 60 F_6 -48 E_5+10 E_2 F_3 +16\zeta(5), \non \\ 10 F_9 &=& 20 F_6 -4 E_5 +3\zeta(5),\eea
and the expressions for $B_1$ and $B_2$ obtained above, we get that
\be \mathcal{A}_{D^8\mathcal{R}^5} = -\frac{24\pi}{5} \int_{\mathcal{F}_L} \frac{d^2\tau}{\tau_2^2} \Big( \frac{13}{4} F_6 +\frac{3}{2} E_5 - E_2 E_3 -\frac{1}{4} E_2 F_3 +\frac{\zeta(5)}{12}\Big) .\ee
Using the Poisson equation~\cite{D'Hoker:2016jac,Basu:2016kli}
\be (\Delta -6) F_6 = \frac{86}{5} E_5 - 4 E_2 E_3 +\frac{\zeta(5)}{10},\ee
and the asymptotic expansions of $F_{6}$ and $E_s$ we evaluate the integral once again along the lines of~\cite{D'Hoker:2015foa}, to obtain
\be  \mathcal{A}_{D^8\mathcal{R}^5} =  -\frac{7\zeta(2)\zeta(5)}{25}.\ee

For the $D^{10}\mathcal{R}^5$ interaction, there are two distinct contributions given by
\bea \mathcal{A}_{D^{10}\mathcal{R}^5}^{(1)} &=&-2\pi\int_{\mathcal{F}_L} \frac{d^2\tau}{\tau_2^2} \Big(-\frac{1}{6} F_{14} +F_{15} +\frac{55}{144} F_{16} +\frac{2}{3} F_{17} -2 F_{18} + F_{19} -\frac{235}{108} F_{20} \non \\ &&-\frac{7}{6} F_{21} -F_{22}-\frac{1}{2}F_{23} + F_{24} -\frac{667}{12960} F_{25} -4 E_2 E_4 - 3 E_3^2 +\frac{7}{6} E_3 F_3 \non \\ &&+\frac{257}{288} E_2 F_{11} -5 E_2 F_{12} +\frac{667}{1296} F_3^2 +\frac{191}{144} E_2^3 -\frac{2}{\pi} B_3 +\frac{4}{\pi} B_4 -\frac{2}{\pi}B_5 \non \\ &&+\frac{2}{\pi} B_6 +\frac{3}{\pi}(B_7 + c.c.) +\frac{4}{\pi} B_8 +\frac{1}{6\pi} B_9\Big), \eea
and
\bea \mathcal{A}_{D^{10}\mathcal{R}^5}^{(2)} &=&-2\pi\int_{\mathcal{F}_L} \frac{d^2\tau}{\tau_2^2} \Big( \frac{3}{18} F_{14} +F_{15} +\frac{355}{576} F_{16}+\frac{1}{3} F_{17} -2F_{18} -2F_{19}-\frac{211}{432} F_{20}\non \\ &&-\frac{1}{2} F_{21} -\frac{1}{8} F_{23} +F_{24} +\frac{E_6}{10368} -E_2 E_4 -\frac{1}{2} E_3^2 +\frac{341}{1152} E_2 F_{11} -\frac{1}{4} E_2 F_{12} \non \\ &&+\frac{1}{2} E_3 F_3 +\frac{1291}{5184} F_3^2 +\frac{101}{576} E_2^3 -\frac{2}{\pi}B_5+\frac{1}{\pi} B_6+\frac{1}{\pi}(B_7 + c.c.) +\frac{1}{\pi}B_8 \Big) .\eea

These can be simplified using the relations~\cite{Basu:2016kli}
\bea 3 F_{22} = F_{17} + 12 F_{18} - 4 E_6,\eea
and~\cite{DHoker:2016quv}
\bea 3 F_{14} &=& 36 F_{15} +109 F_{17} +408 F_{18} -211 E_6 +12 E_3 F_3 + 18 E_2 F_{12} , \non \\ 18 F_{19} &=& 18 F_{15} -\frac{9}{2} F_{16} + 35 F_{17} +132 F_{18} - 50 E_6 +\frac{11}{5880}\zeta(3)^2, \non \\ 2 F_{20} &=& 12 F_{15}+\frac{3}{2} F_{16} -3 F_{17} -12 E_6 + 6 E_2 E_4 +\frac{7}{216}\zeta(3)^2, \non \\ 2 F_{21} &=& 9 F_{16} + 39 F_{17} + 144 F_{18} -63 E_6 + 6 E_2 F_{12} +\frac{7}{36}\zeta(3)^2, \non \\ 3 F_{23} &=& 36 F_{15}-9 F_{16} +88 F_{17} +336 F_{18}-166 E_6+ 12 E_3^2 + 3E_2^3 - \frac{7}{36}\zeta(3)^2,\non \\ 6 F_{24} &=& 12 F_{15}-3 F_{16} + 10 F_{17} + 48 F_{18} -40 E_6 + 6 E_2 E_4 + 12 E_3^2 +\frac{11}{8820}\zeta(3)^2, \non \\ F_{25} &=& -720 F_{15}+360 F_{16} +2260 F_{17}+7680 F_{18}-2800 E_6 +10F_3^2 +15 E_2 F_{11} \non \\ && -30 E_2^3 +\frac{70}{9}\zeta(3)^2,  \eea
as well as the expressions for $B_3, B_4, B_5, B_6, B_7 + c.c., B_8$ and $B_9$ obtained above.
We get that
\bea \mathcal{A}_{D^{10}\mathcal{R}^5}^{(1)} =-\frac{\pi}{324}\int_{\mathcal{F}_L} \frac{d^2\tau}{\tau_2^2} \Big( -34344 F_{15} -3339 F_{16} -16133 F_{17} - 93264 F_{18} +44231 E_6 \non \\ + 6 E_2 (24 E_4 + 85 F_{11} - 540 F_{12}) +81 (4 E_3 + F_3)^2+ 618 E_2^3-\frac{361589}{3528}\zeta(3)^2\Big),\eea
and
\bea \mathcal{A}_{D^{10}\mathcal{R}^5}^{(2)}=-2\pi\int_{\mathcal{F}_L} \frac{d^2\tau}{\tau_2^2} \Big( -\frac{715}{72} F_{15} -\frac{9}{8} F_{16} -\frac{11989}{3744} F_{17} -\frac{91}{3} F_{18} + \frac{118369}{10368} E_6 \non \\ +\frac{E_2}{1152}(1768 E_4 + 149 F_{11} + 576 F_{12} - 134 E_2^2)- \frac{5}{5184} F_3^2 -\frac{737057}{45722880} \zeta(3)^2\Big) .\eea

To evaluate the contributions involving $F_{15}, F_{16}, F_{17}$ and $F_{18}$, we use the Poisson equations~\cite{D'Hoker:2015foa}
\bea (\Delta -2) (F_{17} + 4 F_{18}) &=& 52 E_6 - 4 E_3^2, \non \\ (\Delta -12) (-F_{17} +6 F_{18}) &=& 108 E_6 - 36 E_3^2, \non \\ (\Delta -12)(6 F_{15}+ F_{17}) &=& 120 E_6 + E_3^2 - 36 E_2 E_4\eea
and~\cite{Basu:2016xrt}
\bea (\Delta -2) F_{16} &=& 32 F_{15} -\frac{16}{9} F_{17} +\frac{128}{3}F_{18}-\frac{56}{9} E_6 -12 E_2 E_4 -8 E_2 F_{12} \non \\ &&+ 8 E_3^2 + \frac{4}{3} E_2^3 +\frac{11}{13230}\zeta(3)^2.\eea

Thus dropping terms in the integrand proportional to $E_6, E_3$ and $E_2 E_4$ that do not contribute to the amplitude, we get that
\bea \label{e1}\mathcal{A}_{D^{10}\mathcal{R}^5}^{(1)} =-\frac{\pi}{324} \Big[ -\frac{3339}{2}F'_{16} -\frac{77859}{2} (F_{17}' + 4 F_{18}')+\frac{5591}{10} (-F_{17}' + 6 F_{18}') \non \\ + 265 (6 F_{15}' + F_{17}')  + 4374 I_1 -4356 I_2 + 4100 I_3 -\frac{354619 \pi}{52920}\zeta(3)^2\Big] \eea 
where we have defined the boundary contributions
\be X' \equiv \frac{\p X}{\p \tau_2} \Big\vert_{\tau_2 =  L\rightarrow \infty},\ee
and we have kept only terms that are finite as $L\rightarrow \infty$.
We have also defined the integrals
\bea I_1 = \int_{\mathcal{F}_L} \frac{d^2\tau}{\tau_2^2} E_2^3, \quad I_2 = \int_{\mathcal{F}_L} \frac{d^2\tau}{\tau_2^2}E_2 F_{12} , \quad I_3 = \int_{\mathcal{F}_L} \frac{d^2\tau}{\tau_2^2} E_3^2.\eea
Thus terms of the type $X'$ in \C{e1} contribute only if $X$ contains a term linear in $\tau_2$ in the large $\tau_2$ expansion. These contributions are given by
\be F_{15} = -\frac{\pi \tau_2}{1890} \zeta(5), \quad F_{16} = -\frac{\pi \tau_2}{945} \zeta(5), \quad F_{17} =0, \quad F_{18} =  -\frac{\pi \tau_2}{630} \zeta(5).\ee
where we have dropped all other contributions apart from the term linear in $\tau_2$.
Also $I_3$ can be evaluated to give~\cite{Basu:2015ayg}
\be I_3 = -\frac{\pi \zeta(5)}{315} {\rm ln} \s,\ee
where
\be {\rm ln} \s = \frac{\zeta'(5)}{\zeta(5)} -\frac{\zeta'(6)}{\zeta(6)} +\frac{7}{12} -{\rm ln}2.\ee
Putting these contributions together, we get that
\bea \mathcal{A}_{D^{10}\mathcal{R}^5}^{(1)} = \frac{410}{1701}\zeta(2)\zeta(5) {\rm ln}\hat{\s} +\frac{354619}{2857680} \zeta(2)\zeta(3)^2 - \frac{27\pi}{2} I_1 +\frac{121\pi}{9} I_2,\eea
where 
\be {\rm ln} \hat\s = {\rm ln} \s + \frac{379451}{20500}.\ee
We have not evaluated the integrals $I_1$ and $I_2$. It should perhaps be possible to evaluate them along the lines of~\cite{Zagier}, and it would be interesting to explicitly do so. Thus on using the various relations among the graphs without derivatives, and the relations that express the graphs with derivatives in terms of those without derivatives, we see that the structure of these terms simplify considerably.   

The analysis for $\mathcal{A}_{D^{10}\mathcal{R}^5}^{(2)}$ can be done exactly in the same way, as well as the analysis of the $D^{12} \mathcal{R}^4$ interaction resulting from the low momentum expansion of the four graviton amplitude. 

\appendix

\section{Some formulae involving Green functions}

In the various modular graph functions, the scalar Green function which connect the vertices is given by
~\cite{Green:1999pv,Green:2008uj}
\be \label{Green}G(z;\tau) = \frac{1}{\pi} \sum_{(m,n)\neq(0,0)} \frac{\tau_2}{\vert m\tau+n\vert^2} e^{\pi[\bar{z}(m\tau+n)-z(m\bar\tau+n)]/\tau_2}.\ee
Thus $G(z;\tau)$ is modular invariant, and single valued, hence
\be \label{sv}G(z;\tau) = G(z+1;\tau) = G(z+\tau;\tau).\ee

The Green function satisfies the equations
\bea \label{eigen}\bar{\p}_w\p_z G(z,w) = \pi \delta^2 (z-w) - \frac{\pi}{\tau_2}, \non \\ \bar{\p}_z\p_z G(z,w) = -\pi \delta^2 (z-w) + \frac{\pi}{\tau_2} \eea
which is repeatedly used in our analysis. Here $z$ is the coordinate on the torus given by
\be \label{range}-\frac{1}{2} \leq {\rm Re} z \leq \frac{1}{2}, \quad 0 \leq {\rm Im} z \leq \tau_2 .\ee

In the various manipulations, we often obtain expressions involving $\p_z G(z,w)$ where $z$ is integrated over $\S$, the worldsheet of the torus. We then integrate by parts without picking up boundary contributions on $\S$ as $G(z,w)$ is single valued. Hence we also drop all contributions which are total derivatives as they vanish. Also we readily use $\p_z G(z,w) = -\p_w G(z,w)$ using the translational invariance of the Green function. Finally, we have that
\be \int_\S d^2 z G(z,w)=0\ee
which easily follows from \C{Green}. Hence one particle reducible diagrams vanish.

In the various diagrams, the notations for holomorphic and antiholomorphic derivatives acting on the Green function are given in figure 26, along with the notation for a single Green function having both these derivatives.  

\begin{figure}[ht]
\begin{center}
\[
\mbox{\begin{picture}(220,60)(0,0)
\includegraphics[scale=.7]{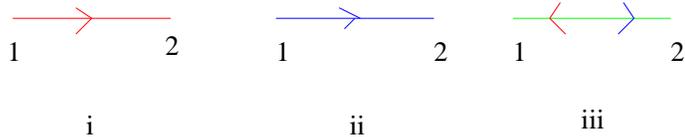}
\end{picture}}
\]
\caption{ (i) $\p_2 G_{12}$, (ii) $\bar\p_2 G_{12}$ and (iii) $\p_1 \bar\p_2 G_{12}$}
\end{center}
\end{figure}

Some of the graphs are expressed in term of $E_s$, the $SL(2,\mathbb{Z})$ invariant non--holomorphic Eisenstein series defined by
\be \label{Eisenstein}E_s (\tau,\bar\tau) = \sum_{(m,n)\neq (0,0)}\frac{\tau_2^s}{\pi^s\vert m+n\tau\vert^{2s}}.\ee        

In the expressions for the various graphs, the integral over the position of insertion of the vertex operator is given by
\be \frac{1}{\tau_2}\int_z \equiv \int_\S \frac{d{\rm Re}z d{\rm Im}z}{\tau_2}\ee
where $\S$ is the worldsheet of the torus, and $z$ is the coordinate satisfying \C{range}, and all vertices are integrated over. For example in figure 4, we have that
\be B_8 = \frac{1}{\tau_2^4} \prod_{i=1}^4 \int_{z_i} \p_{z_2} G (z_{12}) G(z_{23}) G(z_{34}) G(z_{14}) G(z_{13})^2 \bar\p_{z_3} G(z_{13}),\ee 
where
\be G(z_{ij}) \equiv G(z_i - z_j;\tau).\ee
For the various auxiliary diagrams, the number of factors of $\tau_2^{-1}$ in the integrals are determined in the obvious way such that the various relations are satisfied. For example in figure 22, we have that
\be F_{50} = \frac{1}{\tau_2^4}\prod_{i=1}^5 \int_{z_i} G(z_{12}) \p_{z_3} G(Z_{23}) \bar\p_{z_3} G(z_{34}) G(z_{14}) G(Z_{35})^2 \p_{z_3}G(z_{35}) \bar\p_{z_1} G(z_{15}).\ee  

%\bibliographystyle{utphys}
%\bibliography{myrefs}

\providecommand{\href}[2]{#2}\begingroup\raggedright\endgroup

\end{document}